\documentclass[useAMS,usenatbib]{mn2e}
\usepackage[dvips]{graphicx}
\usepackage{epsfig,amssymb,amsmath,color}
\usepackage{algorithm,algorithmic,multirow,hhline}
\newcommand{\beq}{\begin{equation}}
\newcommand{\eeq}{\end{equation}}
\newcommand{\beqn}{\begin{eqnarray}}
\newcommand{\eeqn}{\end{eqnarray}}

\def\bmath#1{\mbox{\boldmath$#1$}}

\long\def\symbolfootnote[#1]#2{\begingroup%
\def\thefootnote{\fnsymbol{footnote}}\footnote[#1]{#2}\endgroup}

\title[Robust Radio Interferometric Calibration Using the t-Distribution]{Robust Radio Interferometric Calibration Using the t-Distribution}
\author[Kazemi and Yatawatta]{S. Kazemi$^{1}$\thanks{E-mail:
kazemi@astro.rug.nl} and S. Yatawatta$^{2}$\\
$^{1}$Kapteyn Astronomical Institute, University
of Groningen, P.O. Box 800, 9700 AV Groningen, the Netherlands\\
$^{2}$ASTRON, Postbus 2, 7990 AA Dwingeloo, the Netherlands}
\begin{document}
\date{\today}
\pagerange{\pageref{firstpage}--\pageref{lastpage}} \pubyear{2012}
\maketitle
\label{firstpage}
%

\begin{abstract}
A major stage of radio interferometric data processing is calibration or the estimation of systematic errors in the data and the correction for such errors. A stochastic error (noise) model is assumed, and in most cases, this underlying model is assumed to be Gaussian. However, outliers in the data due to interference or due to errors in the sky model would have adverse effects  on processing based on a Gaussian noise model. Most of the shortcomings of calibration such as the loss in flux or coherence, and the appearance of spurious sources, could be attributed to the deviations of the underlying noise model. In this paper, we propose to improve the robustness of calibration by using a noise model based on Student's t distribution. Student's t noise is a special case of Gaussian noise when the variance is unknown. Unlike Gaussian noise model based calibration, traditional least squares minimization would not directly extend to a case when we have a Student's t noise model. Therefore, we use a variant of the Expectation Maximization (EM) algorithm, called the Expectation-Conditional Maximization Either (ECME) algorithm  when we have a Student's t noise model and use the Levenberg-Marquardt algorithm in the maximization step. We give simulation results to show the robustness of the proposed calibration method as opposed to traditional Gaussian noise model based calibration, especially in preserving the flux of weaker sources that are not included in the calibration model.
\end{abstract}

\section{Introduction}
Radio interferometry gives an enhanced view of the sky, with increased sensitivity and higher resolution. There is a trend towards using phased arrays as the building blocks of radio telescopes (LOFAR \footnote{The Low Frequency Array}, SKA \footnote{The Square Kilometer Array}) as opposed to traditional dish based interferometers. In order to reach the true potential of such telescopes, calibration is essential. Calibration refers to estimation of systematic errors introduced by the instrument (such as the beam shape and receiver gain) and also by the propagation path (such as the ionosphere), and correction for such errors, before any imaging is done. Conventionally, calibration is done by observing a known celestial object (called the external calibrator), in addition to the part of the sky being observed. This is improved by self-calibration \citep{Cornwell0}, which is essentially using the observed sky  itself for the calibration. Therefore, self calibration entails consideration of  both the sky as well as the instrument as unknowns. By iteratively refining the sky and the instrument model, the quality of the calibration is improved by orders of magnitude in comparison to using an external calibrator.

 From a signal processing perspective, calibration is essentially the Maximum Likelihood (ML) estimation of the instrument and sky parameters. An in depth overview of existing calibration techniques from an estimation perspective can be found in e.g. \cite{Boonstra03},\cite{AJV04},\cite{Jeffs06} and \cite{Kaz2}. All such calibration techniques are based on a Gaussian noise model and the ML estimate is obtained by minimizing the least squares cost function using a nonlinear optimization technique such as the Levenberg-Marquardt (\cite{Lev44},\cite{Mar63}) (LM) algorithm. Despite the obvious advantages of self-calibration, there are also some limitations. For instance, \cite{Cornwell} give a detailed overview of the practical problems in self-calibration, in particular due to errors in the initial sky model. It is well known that the sources not included in the sky model have lower flux (or loss of coherence) and \cite{vidal10} is a recent study on this topic. Moreover, under certain situations, fake or spurious sources could appear due to calibration as studied by \cite{vidal08}.

In this paper, we propose to improve the robustness of calibration by assuming a Student's t \citep{Student} noise model instead of a Gaussian noise model. One of the earliest attempts in deviating from a Gaussian noise model based calibration can be found in \cite{Schwab82}, where instead of minimizing a least squares cost function, an $l_1$ norm minimization was considered. Minimizing the $l_1$ norm is equivalent to having a noise model which has a  Laplacian distribution \citep{Aravkin}. The motivation for \cite{Schwab82} to deviate from the Gaussian noise model was the ever present outliers in the radio interferometric data.

In a typical  radio interferometric observation, there is a multitude of causes for outliers in the data:
\begin{itemize}
\item Radio frequency interference caused by man made radio signals is a persistent cause of outliers in the data. However, data affected by such interference is removed before any calibration is performed by flagging (e.g. \cite{aoflagger}). But there might be faint interference still present in the data, even after flagging. 
\item The initial sky model used in self calibration is almost always different from the true sky that is observed. Such model errors also create outliers in the data. This is especially significant when we observe a part of the sky that has sources with complicated, extended structure. Moreover, during calibration, only the brightest sources are normally included in the sky model and the weaker sources act together to create outliers.
\item During day time observations, the Sun could act as a source of interference, especially during high solar activity. In addition the Galactic plane also affects the signals on short baselines.
\item An interferometer made of phased arrays will have sidelobes that change with time and frequency. It is possible that a strong celestial source, far away from the part of the sky being observed, will pass through such a sidelobe. This will also create outliers in the data.
\end{itemize} 
To summarize, model errors of the sky as well as the instrument will create outliers in the data and in some situations calibration based on a Gaussian noise model will fail to perform satisfactorily. In this paper, we consider the specific problem of the effect of unmodeled sources in the sky during calibration. We consider 'robustness' to be the preservation of the fluxes of the unmodeled sources. Therefore, our prime focus is to minimize the loss of flux or coherence of unmodeled sources and our previous work \citep{Vietnam} have measured robustness in terms of the quality of calibration.

Robust data modeling using Student's t distribution has been applied in many diverse areas of research and \cite{Lange}, \cite{Bart} and \cite{Aravkin} are few such examples. However, the traditional least squares minimization is not directly applicable when we have a non-Gaussian noise model, and we apply the  Expectation Maximization (EM) \citep{DLR} algorithm to convert calibration into an iteratively re-weighted least squares minimization problem, as proposed by \cite{Lange}. In fact, we use an extension of the EM algorithm called the Expectation-Conditional Maximization Either (ECME) algorithm \citep{Liu1995} to convert calibration to a tractable minimization problem. However, we emphasize that we do not force a non-Gaussian noise model onto the data. In the case if there are no outliers and the noise is actually Gaussian, our algorithms would work as traditional calibration would.

The rest of the paper is organized as follows: In section \ref{model}, we give an overview of radio interferometric calibration. We consider the effect of unmodeled sources in the sky during calibration in section \ref{noiseprop}. Next, in section \ref{robustcal}, we discuss the application of Student's t noise model in calibration. We also present the weighted LM routine used in calibration. In section \ref{simulations}, we present simulation results to show the superiority of the proposed calibration approach in minimizing the loss in coherence and present conclusions in section \ref{conclusions}.

{\em Notation}: Lower case bold letters refer to column vectors (e.g. ${\bmath y}$). Upper case bold letters refer to matrices (e.g. ${\bf {\sf C}}$). Unless otherwise stated, all parameters are complex numbers. The matrix inverse, transpose, Hermitian transpose, and conjugation are referred to as $(.)^{-1}$, $(.)^{T}$, $(.)^{H}$, $(.)^{\star}$, respectively. The matrix Kronecker product is given by $\otimes$. The statistical expectation operator is given as $E\{.\}$. The vectorized representation of a matrix is given by $\mathrm{vec}(.)$. The $i$-th diagonal element of matrix ${\bf {\sf A}}$ is given by ${\bf {\sf A}}_{ii}$. The $i$-th element of a vector ${\bmath y}$ is given by ${\bmath y}_i$. The identity matrix is given by ${\bf {\sf I}}$.  Estimated parameters are denoted by a hat, $\widehat{(.)}$. All logarithms are to the base $e$, unless stated otherwise. The $l_2$ norm is given by $\|.\|$ and the infinity norm is given by $\|.\|_{\infty}$. A random variable $X$ that has a distribution $\mathcal P$ is denoted by $X \sim {\mathcal P}$.

\section{Data Model}\label{model}
We give a brief overview of radio interferometry in this section.  For more information about radio interferometry, the reader is referred to \cite{TMS}, and \cite{HBS} for the data model in particular. We consider the radio frequency sky to be composed of discrete sources, far away from the earth such that the approaching radiation from each one of them appears to be plane waves.
 We decompose the contribution from the $i$-th source into two orthogonal polarizations ${\bmath u}_i=[u_{xi}\ u_{yi}]^T$. The interferometric array consists of $R$ receiving elements with dual polarized feeds. At the $p$-th station, this plane wave causes an induced voltage, which is dependent on the beam attenuation as well as the radio frequency receiver chain attenuation. The induced voltages at the $x$ and $y$ polarization feeds, $\tilde{\bmath v}_{pi}=[v_{xpi}\ v_{ypi}]^T$  due to source $i$ are given as
\beq \label{ind}
\tilde{\bmath v}_{pi}={\bf {\sf J}}_{pi} {\bmath u}_i.
\eeq
The $2$ by $2$ Jones matrix ${\bf {\sf J}}_{pi}$ in  (\ref{ind}) represents the effects of the propagation medium, the beam shape and the receiver. If there are $K$ known sources (that are in the sky model) and $K^{\prime}$ unknown sources, the total signal will be a superposition of $K+K^{\prime}$ such signals as in (\ref{ind}). 
\begin{figure}
\begin{minipage}{0.99\linewidth}
\centering
\input{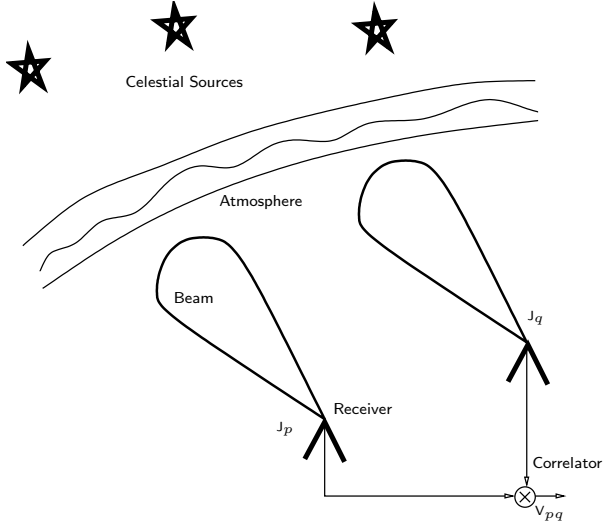}
\end{minipage}
\caption{A basic radio interferometer that correlates the signals received from far away celestial sources. The signals are corrupted by the earth's atmosphere as well as by the receiver beam pattern, and these corruptions are represented by ${\bf {\sf J}}_p$ and ${\bf {\sf J}}_q$.\label{R}}
\end{figure}

Consider the correlation of signals at the $p$-th receiver and the $q$-th receiver, as shown in Fig. \ref{R}, with proper signal delay. After correlation, the correlated signal of the $p$-th station and the $q$-th station (named as the {\em visibilities}), ${\bf {\sf V}}_{pq}=E\{ {\bmath v}_{p} {\bmath v}_{q}^H \}$ is given by 
\beq \label{vispq}
{\bf {\sf V}}_{pq}= \sum_{i=1}^{K}  {\bf {\sf J}}_{pi} {\bf {\sf C}}_{pqi} {\bf {\sf J}}_{qi}^{H} + \sum_{i^{\prime}=1}^{K^{\prime}}  {\bf {\sf J}}_{pi^{\prime}} {\bf {\sf C}}_{pqi^{\prime}} {\bf {\sf J}}_{qi^{\prime}}^{H} + {\bf {\sf N}}_{pq}.
\eeq
In (\ref{vispq}), ${\bf {\sf J}}_{pi}$ and ${\bf {\sf J}}_{qi}$ are the Jones matrices describing errors along the direction of source $i$, at station  $p$ and $q$, respectively. The $2$ by $2$ noise matrix is given as ${\bf {\sf N}}_{pq}$. The contribution from the $i$-th source on baseline $pq$ is given by the $2$ by $2$ matrix ${\bf {\sf C}}_{pqi}$. The noise matrix ${\bf {\sf N}}_{pq}$ is assumed to have elements with zero mean, complex Gaussian entries with equal variance in real and imaginary parts. Moreover, in (\ref{vispq}), we have split the contribution from the sky into two parts: $K$ sources that are known to us and $K^{\prime}$ sources that are unknown. Generally, the bright sources are always known but there are infinitely many faint sources that are too weak to be detected and too numerous to be included in the sky model. Therefore, almost always $K^{\prime}$ is much larger than $K$.

During calibration, we only estimate the Jones matrices ${\bf {\sf J}}_{pi}$ for $p\in[1,R]$ and $i\in[1,K]$, in other words, we estimate the errors along the known bright sources. Due to our ignorance of the $K^{\prime}$ unknown sources, the effective noise during calibration becomes
\beq \label{nblock}
{\bf {\sf N}}^{\prime}_{pq}= \sum_{i^{\prime}=1}^{K^{\prime}}  {\bf {\sf J}}_{pi^{\prime}} {\bf {\sf C}}_{pqi^{\prime}} {\bf {\sf J}}_{qi^{\prime}}^{H} + {\bf {\sf N}}_{pq}
\eeq 
and our assumption regarding the noise being complex circular Gaussian breaks down, depending on the properties of the signals of the weak sources. The prime motivation of this paper is to address this problem of the possible non-Gaussianity of the noise due to an error in the sky model. A similar situation could arise even for calibration along one direction (or direction independent calibration), when $K=1$, if there is an error in the source model, for instance in the shape of the source.

The vectorized form of (\ref{vispq}), ${\bmath v}_{pq}=\mathrm{vec}({\bf {\sf V}}_{pq})$  can be written as 
\beq \label{vecvispq}
{\bmath v}_{pq}= \sum_{i=1}^K {\bf {\sf J}}_{qi}^{\star}\otimes {\bf {\sf J}}_{pi} \mathrm{vec}({\bf {\sf C}}_{pqi}) +  \sum_{i^{\prime}=1}^{K^{\prime}}{\bf {\sf J}}_{qi^{\prime}}^{\star}\otimes {\bf {\sf J}}_{pi^{\prime}} \mathrm{vec}({\bf {\sf C}}_{pqi^{\prime}})  + {\bmath n}_{pq}
\eeq
  where ${\bmath n}_{pq}=\mathrm{vec}({\bf {\sf N}}_{pq})$. Depending on the time and frequency interval within which calibration solutions are obtained, we can stack up all cross correlations within that interval as
\beq
{\bmath d}=[\mathrm{real}({\bmath v}^T_{12})\ \mathrm{imag}({\bmath v}^T_{12})\ \mathrm{real}({\bmath v}^T_{13})\ldots \ldots \mathrm{imag}({\bmath v}^T_{(R-1)R})]^T
\eeq
where  ${\bmath d}$ is a vector of size $N\times 1$ of real data points. Thereafter, we can rewrite the data model as
\beq \label{obs}
{\bmath d}=\sum_{i=1}^{K}{\bmath s}_i({\bmath \theta}) +\sum_{i^{\prime}=1}^{K^{\prime}} {\bmath s}_{i^{\prime}} + {\bmath n}
\eeq
where ${\bmath \theta}$ is the real parameter vector (size $M\times 1$) that is estimated by calibration. The contribution of the $i$-th known source on all data points is given by ${\bmath s}_i({\bmath \theta})$ (size $N\times 1$) and the unknown contribution from the $i^{\prime}$-th unknown source is given by ${\bmath s}_{i^{\prime}}$ (size $N\times 1$). The noise vector based on a Gaussian noise model is given by  ${\bmath n}$ (size $N\times 1$). The parameters ${\bmath \theta}$ are the elements of ${\bf {\sf J}}_{pi}$-s, with real and imaginary parts considered separately. 

The ML estimate of ${\bmath \theta}$ under a zero mean, white Gaussian noise is obtained by minimizing the least squares cost 
\beq \label{mltheta}
\widehat{\bmath \theta}=\underset{\bmath \theta}{\rm arg\ min}\|{\bmath d}- \sum_{i=1}^{K}{\bmath s}_i({\bmath \theta})\|^2
\eeq
as done in current calibration approaches (\cite{Boonstra03},\cite{AJV04},\cite{Jeffs06},\cite{Kaz2}).  However, due to the unmodeled sources, the effective noise is actually
\beq \label{effectivenoise}
{\bmath n}^{\prime}=\sum_{i^{\prime}=1}^{K^{\prime}} {\bmath s}_{i^{\prime}} + {\bmath n}
\eeq
even when ${\bmath n}$ is assumed to be Gaussian. Therefore, traditional calibration based on a least squares cost minimization would not perform optimally. In order to improve this, we have to consider the statistical properties of the effective noise ${\bmath n}^{\prime}$ and we shall do that in the section \ref{noiseprop}.  
\section{Effect of unmodeled sources in calibration}\label{noiseprop}
In this section we study the effect of unmodeled sources on ${\bf {\sf N}}^{\prime}_{pq}$ in (\ref{nblock}) when ${\bf {\sf N}}_{pq}$ has elements with zero mean, complex circular white Gaussian statistics. We only select one element from the $2\times 2$ matrix (say at $1$-st row and column) for simplicity. Let us denote the baseline coordinates as $u,v,w$ in wavelengths (we omit the $pq$ subscript for simplicity). We can rewrite (\ref{nblock}) for just one element as
\beq \label{scalarn}
z_{pq}=\sum_{i^{\prime}=1}^{K^{\prime}} g_{pq i^{\prime}} I_{pq i^{\prime}} \exp \left( -\jmath 2\pi \left( u l_{i^{\prime}} + v m_{i^{\prime}} + w (n_{i^{\prime}}-1)\right) \right) + n_{pq}.
\eeq
In (\ref{scalarn}), $g_{pq i^{\prime}}$ correspond to the corruptions along the direction $i^{\prime}$ (contributions from ${\bf {\sf J}}_{pi^{\prime}}$  and ${\bf {\sf J}}_{qi^{\prime}}$). The intensity of the $i^{\prime}$-th source seen on baseline $pq$ is given by $I_{pq i^{\prime}}$. The direction cosines of the $i^{\prime}$-th source are given by $l_{i^{\prime}},m_{i^{\prime}},n_{i^{\prime}}$. The Gaussian noise is given by $n_{pq} \sim \mathcal{CN}(0,\rho^2)$. We assume that $g_{pq i^{\prime}}$, $I_{pq i^{\prime}}$, $l_{i^{\prime}},m_{i^{\prime}},n_{i^{\prime}}$ and $n_{pq}$ are statistically independent from each other. Moreover, the sources are assumed to be uniformly distributed in a field of view defined by $-\overline{l} \le l_{i^{\prime}} \le \overline{l}$ and $-\overline{m} \le m_{i^{\prime}} \le \overline{m}$ and that $(n_{i^{\prime}}-1) \approx 0$. The sources outside this area in the sky have almost no contribution to the signal due to the fact that the values of $|g_{pq i^{\prime}}|$ are very small, mainly due to attenuation by the beam shape.

With the above assumptions, we see that
\beq  \label{scalarne}
E\{z_{pq}\}=\sum_{i^{\prime}=1}^{K^{\prime}} E\{g_{pq i^{\prime}}\} E\{I_{pq i^{\prime}}\} \mathrm{sinc}\left(2\pi u \overline{l}\right) \mathrm{sinc}\left(2\pi v \overline{m}\right) + E\{n_{pq}\}
\eeq
which is almost zero if $|u|>\frac{1}{2  \overline{l}}$ and $|v|>\frac{1}{2  \overline{m}}$. Therefore, we make the following assumptions applicable to long baselines:
\begin{itemize}
\item The mean of the effective noise is almost equal to the mean of noise, $E\{z_{pq}\} \rightarrow E\{n_{pq}\} = 0$.
\item The variance of effective noise is greater than the variance of noise, $E\{|z_{pq}|^2\} > E\{|n_{pq}|^2\}$.
\end{itemize}

Let us briefly consider the implications of (\ref{scalarne}) above. First, the field of view is $2\overline{l} \times 2\overline{m}$ in the sky. Now, consider the longest baseline length or the maximum value of $\sqrt{u^2+v^2}$ to be $\overline{u}$. Therefore, the image resolution will be about $1/\overline{u}$ and consider the field of view to be of width $ 2 \overline{l} \approx P\times 1/\overline{u}$. In other words, the field of view is $P$ image pixels when the pixel width is $1/\overline{u}$. Now, in order for $E\{z_{pq}\} \approx 0$ in (\ref{scalarne}),  we need $|u|>\frac{1}{2  \overline{l}}$, or $|u|> \overline{u}/P$ (and a similar expression can be derived for $|v|$). This means that for baselines that are at least $1/P$ times the maximum baseline length, we can assume $E\{z_{pq}\} \approx 0$.

To illustrate the above discussion, we give a numerical example considering the LOFAR highband array at 150 MHz. The point spread function at this frequency is about $6^{\prime\prime}$ and the field of view is about 10 degrees in diameter. Therefore, $P\approx 10\times3600/6 = 6000$. The longest baselines is about 80 km and for all baselines that are greater than $80/6=13$ m, the assumptions made above more or less hold.

To summarize the discussion in this section, we claim that
\beq
E\{\sum_{i^{\prime}=1}^{K^{\prime}} {\bmath s}_{i^{\prime}}\} \rightarrow {\bmath 0}
\eeq
in (\ref{effectivenoise}) and therefore, $E\{{\bmath n}^{\prime}\}\rightarrow  E\{{\bmath n}\}$. However, the covariance of ${\bmath n}^{\prime}$ is different than the covariance of ${{\bmath n}}$ and in general, the effective noise is not necessarily Gaussian anymore.

\subsection{SAGE algorithm with unmodeled sources}\label{sec:sage}
In our previous work \citep{Kaz2}, we have presented the Space Alternating Generalized Expectation Maximization (SAGE) \citep{Fess94} algorithm as an efficient and accurate method to solve (\ref{mltheta}), when the noise model is Gaussian. However, when there are unmodeled sources, as we have seen in this section, the noise model is not necessarily Gaussian. 

The SAGE Expectation step is finding the conditional mean of the $k$-th signal,
\beq
{\bmath x}^k= {\bmath s}_k({\bmath \theta}) + {\bmath n}^{\prime}={\bmath s}_k({\bmath \theta}) + \sum_{i^{\prime}=1}^{K^{\prime}} {\bmath s}_{i^{\prime}} + {\bmath n}
\eeq
where ${\bmath x}^k$ is the hidden data. Using this, we can rewrite the observed data ${\bmath d}$ as 
\beq
{\bmath d}={\bmath x}^k + \sum_{i=1,i\ne k}^{K}{\bmath s}_i({\bmath \theta}).
\eeq

The conditional mean of ${\bmath x}^k$ given ${\bmath d}$, is given as $\widehat{{\bmath x}^k}$ where
\beq\label{condmean}
\widehat{{\bmath x}^k}={\bmath s}_k({\bmath \theta}) + \left({\bmath d}-\sum_{i=1,i\ne k}^{K}{\bmath s}_i({\bmath \theta}) -  \sum_{i^{\prime}=1}^{K^{\prime}} {\bmath s}_{i^{\prime}}\right)
\eeq
where we still assume a Gaussian noise model ${\bmath n}$.
Under assumption $\sum_{i^{\prime}=1}^{K^{\prime}} {\bmath s}_{i^{\prime}} \rightarrow {\bmath 0}$, the conditional mean simplifies to
\beq \label{condmean1}
\widehat{{\bmath x}^k}\approx {\bmath s}_k({\bmath \theta}) + \left({\bmath d}-\sum_{i=1,i\ne k}^{K}{\bmath s}_i({\bmath \theta}) \right).
\eeq

The SAGE Maximization step maximizes the likelihood of the conditional mean $\widehat{{\bmath x}^k}$ under the noise ${\bmath n}^{\prime}$. However, we cannot use a least squares cost function as ${\bmath n}^{\prime}$ is not necessarily Gaussian anymore, because of the unmodeled sources. In section \ref{robustcal}, we explore an alternative noise model based on Student's t  distribution \citep{Student} for the maximization of the likelihood.
\section{Robust Calibration}\label{robustcal}
\begin{figure}
\begin{minipage}{0.98\linewidth}
\centering
 \centerline{\epsfig{figure=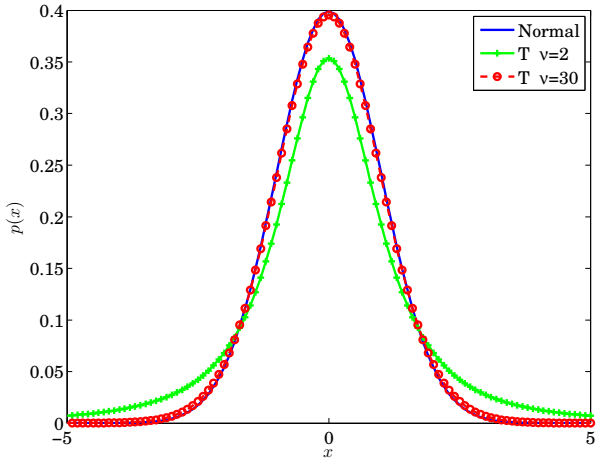,width=8.0cm}}
\end{minipage}
\caption{Probability density functions for standard normal distribution and Student's t distribution, with $\nu=2$ and $\nu=30$. At $\nu=30$, the Student's distribution is indistinguishable from the normal distribution.\label{dist}}
\end{figure}

First, we briefly describe the univariate Student's t distribution  (\cite{Lange}, \cite{Bart}). Let $X$ be a random variable with a normal distribution $\mathcal{N}(\varepsilon,\sigma^2/\gamma)$ where $\gamma$ is also a random variable. Then the conditional distribution of $X$ is
\beq \label{normal_eta}
p(x | \varepsilon,\sigma^2,\gamma)=\frac{1}{(\sigma/\sqrt{\gamma}) \sqrt{2\pi}} \exp{\left(-\frac{1}{2} \left(\frac{x-\varepsilon}{\sigma/\sqrt{\gamma}}\right)^2\right)}.
\eeq
We consider $\gamma$ to have a Gamma distribution, $\gamma \sim {\mathrm {Gamma}}(\frac{\nu}{2},\frac{\nu}{2})$, where $\nu$ is positive (also called the number of degrees of freedom). The density function of $\gamma$ can be given as
\beq
p(\gamma|\nu)=\frac{1}{\Gamma(\frac{\nu}{2})} \left(\frac{\nu}{2}\right)^{\frac{\nu}{2}} \gamma^{\frac{\nu}{2}-1} \exp\left(\frac{-\nu \gamma}{2}\right).
\eeq
Then, the marginal distribution of $X$ is
\beq \label{St}
p(x;\varepsilon,\sigma^2,\nu) =\frac{\Gamma(\frac{\nu+1}{2})}{(\pi \nu)^{1/2} \Gamma(\frac{\nu}{2}) \sigma} \left( 1+ \frac{1}{\nu}\left(\frac{x-\varepsilon}{\sigma}\right)^2\right)^{-\frac{1}{2}(\nu+1)}
\eeq
and this is the probability density function which defines the Student's t distribution.
In Fig. \ref{dist}, we have shown the probability density functions for Gaussian distribution and Student's t distribution, both with zero mean and unit variance. We see that for low values of the number of degrees of freedom $\nu$, Student's t distribution has a higher tail. The asymptotic limit of Student's t distribution is Gaussian as $\nu \rightarrow \infty$, and for $\nu>30$, the two distributions are indistinguishable, within the resolution of the data points used in Fig. \ref{dist}.

Reverting back to (\ref{effectivenoise}), we see that the increase in the noise variance due to the unmodeled sources can be considered as the effect of $\gamma$ in (\ref{normal_eta}). Therefore, we consider the noise vector ${\bmath n}^{\prime}$ to have independent, identically distributed entries, with the distribution given by (\ref{St}) with $\varepsilon=0$ and $\sigma=\rho=1$. In the SAGE iterations outlined in section  \ref{sec:sage}, at the $k$-th iteration (\ref{condmean1}), we have $\widehat{{\bmath x}^k}$ as the data vector and ${\bmath s}_k({\bmath \theta})$ as the model that is used to estimate the parameters ${\bmath \theta}$ (or a subset of the parameters). Therefore, the estimation problem is to find the ML estimate of ${\bmath \theta}$ (size $M\times 1$), given the data ${\bmath y}=\widehat{{\bmath x}^k}$ (size $N\times 1$) and the model ${\bmath f}({\bmath \theta})={\bmath s}_k({\bmath \theta})$ (size $N\times 1$) with noise ${\bmath n}^{\prime}$. Hence, we can rewrite our data model as
\beq
{\bmath y}={\bmath f}({\bmath \theta}) + {\bmath n}^{\prime}
\eeq
where the unknowns are ${\bmath \theta}$ and $\nu$, the number of degrees of freedom of noise ${\bmath n}^{\prime}$.
  Then, the $i$-th element of the vector ${\bmath y}$ (denoted by ${\bmath y}_i$) in (\ref{obs}) will have a similar distribution as (\ref{St}) with $\sigma=1$ and $\mu_i={\bmath f}_i({\bmath \theta})$, where ${\bmath f}_i({\bmath \theta})$ is the $i$-th element of the vector function ${\bmath f}({\bmath \theta})$.
The likelihood function becomes
\beq \label{likelihood}
l({\bmath \theta},\nu|{\bmath y})= \prod_{i=1}^{N} \frac{\Gamma(\frac{\nu+1}{2})}{(\pi \nu)^{1/2} \Gamma(\frac{\nu}{2})} \left( 1+ \frac{\left({\bmath y}_i-{\bmath f}_i({\bmath \theta})\right)^2}{\nu}\right)^{-\frac{1}{2}(\nu+1)}
\eeq
and the log-likelihood function is
\beqn \label{loglikelihood}
\lefteqn{L({\bmath \theta},\nu|{\bmath y})}&&\\\nonumber
&\mbox{} =& N \log \Gamma(\frac{\nu+1}{2}) - N\log \Gamma(\frac{\nu}{2})-\frac{N}{2} \log(\pi\nu)\\\nonumber
&\mbox{} -& \frac{(\nu+1)}{2} \sum_{i=1}^N \log \left(1 +  \frac{({\bmath y}_i-{\bmath f}_i({\bmath \theta}))^2}{\nu} \right).
\eeqn

Note that unlike for the Gaussian case, minimizing a least squares cost function (or maximizing the likelihood) will not give us the ML estimate. In addition, we have an extra parameter, $\nu$, which is the number of degrees of freedom. Hence, we use the Expectation-Conditional Maximization Either algorithm (\cite{Liu1995},\cite{Li2006}) to solve this problem. The ECME algorithm is an extension of the EM algorithm for t distribution presented by \cite{Lange}. 

The auxiliary variables are the weights $w_i$ ($N$ values) and a scalar $\lambda$. All these are initialized to $1$ at the beginning.
The Expectation step in the ECME algorithm involves the conditional estimation of hidden variables $\gamma_i$ (or the weights $w_i$) as
\beq
w_i \leftarrow E\{\gamma_i|{\bmath y}_i, {\bmath \theta}, \nu\} = \lambda \frac{\nu+1}{\nu+({\bmath y}_i-{\bmath f}_i({\bmath \theta}))^2}
\eeq
and the update of the scalar $\lambda$
\beq
\lambda \leftarrow \frac{1}{N} \sum_{i=1}^N w_i.
\eeq
The Maximization step involves finding the value for $\nu$ that is a solution for
\beqn \label{nuzero}
\lefteqn{\Psi(\frac{\nu+1}{2}) - \log\left(\frac{\nu+1}{2}\right) - \Psi(\nu/2) +\log(\nu/2)}\\\nonumber
 &&+ \frac{1}{N}\sum_{i=1}^N\left( \log(w_i)-w_i\right) +1 =0
\eeqn
where  $\Psi(x)=\frac{d}{dx} \log\left(\Gamma(x)\right)$ is the digamma function. Since we know that beyond $\nu>30$, we almost get a Gaussian distribution, and therefore the search space for finding a solution for (\ref{nuzero}) is kept within $2\le \nu \le 30$ and initial value for $\nu$ is chosen to be $2$.

Once $w_i$ is known, ${\bmath y}_i$ has a normal distribution with variance determined by $w_i$. Therefore, in the Maximization step of the EM algorithm, we minimize the weighted least squares cost function 
\beq
l({\bmath \theta}|\nu) =\sum_{i=1}^N w_i ({\bmath y}_i-{\bmath f}_i({\bmath \theta}))^2.
\eeq

 With this formulation at hand, we present the LM algorithm for robust calibration in Algorithm \ref{algLM}, similar to the presentations in \cite{levmar} and \cite{LM}. The additional information needed in Algorithm \ref{algLM} is the Jacobian of ${\bmath f}({\bmath \theta})$, i.e., ${\bf {\sf J}}({\bmath \theta}) = \frac{\partial {\bmath f}({\bmath \theta})}{\partial {\bmath \theta}}$, that can be calculated in closed form using (\ref{vispq}) and (\ref{vecvispq}). The diagonal matrix with the weights $\sqrt{w_i}$ as its diagonal entries is given by ${\bf {\sf W}}$. 

\begin{algorithm}
\caption{Robust Levenberg-Marquardt (ECME)}
\label{algLM}
\begin{algorithmic}[1]
\REQUIRE Data ${\bmath y}$, mapping ${\bmath f}({\bmath \theta})$, Jacobian ${\bf {\sf J}}({\bmath \theta})$, $\nu$, initial value ${\bmath \theta}^0$
\STATE ${\bmath \theta} \gets {\bmath \theta}^0$; $w_i\gets 1$; $\lambda\gets 1$
\WHILE { $l<$ max EM iterations }
\STATE $k \gets 0; \eta \gets 2$
\STATE ${\bf {\sf J}}({\bmath \theta})  \gets {\bf {\sf W}} {\bf {\sf J}}({\bmath \theta})$
\STATE ${\bf {\sf A}} \gets  {\bf {\sf J}}({\bmath \theta})^T{\bf {\sf J}}({\bmath \theta}); {\bmath e}\gets {\bf {\sf W}}({\bmath y}-{\bmath f}({\bmath \theta})); {\bmath g} \gets  {\bf {\sf J}}({\bmath \theta})^T {\bmath e}$
\STATE found $\gets  (\|{\bmath g}\|_{\infty} <\epsilon_1); \mu \gets \tau \max {\bf {\sf A}}_{ii}$
\WHILE {(not found) and ($k<$ max iterations)}
\STATE $k \gets  k+1$; Solve $({\bf {\sf A}} + \mu {\bf {\sf I}}){\bmath h}={\bmath g}$
\IF{$\|{\bmath h}\|<\epsilon_2(\|{\bmath \theta}\|+\epsilon_2)$}
\STATE found $\gets true$
\ELSE
\STATE ${\bmath \theta}_{new} \gets {\bmath \theta} + {\bmath h}$
\STATE $\rho \gets (\|{\bmath e}\|^2 - \|{\bf {\sf W}}({\bmath y}-{\bmath f}({\bmath \theta}_{new}))\|^2)/({\bmath h}^T(\mu {\bmath h}+{\bmath g}))$
\IF{$\rho  > 0$} 
\STATE ${\bmath \theta} \gets  {\bmath \theta}_{new}$
\STATE ${\bf {\sf J}}({\bmath \theta})  \gets {\bf {\sf W}} {\bf {\sf J}}({\bmath \theta})$
\STATE ${\bf {\sf A}} \gets  {\bf {\sf J}}({\bmath \theta})^T{\bf {\sf J}}({\bmath \theta}); {\bmath e}\gets {\bf {\sf W}}({\bmath y}-{\bmath f}({\bmath \theta})); {\bmath g} \gets  {\bf {\sf J}}({\bmath \theta})^T {\bmath e}$
\STATE found $\gets (\|{\bmath g}||_{\infty}\le \epsilon_1)$
\STATE $\mu \gets \mu  \max(1/3,1-(2\rho-1)^3);  \eta\gets 2$
\ELSE
\STATE $\mu \gets \mu \eta;  \eta \gets  2 \eta$
\ENDIF
\ENDIF
\ENDWHILE
\STATE Update weights $w_i\gets \lambda \frac{\nu+1}{\nu + ({\bmath y}_i - {\bmath f}_i({\bmath \theta}))^2}$
\STATE Update $\lambda \gets \frac{1}{N} \sum_{i=1}^N w_i$
\STATE Update $\nu$ using (\ref{nuzero})
\STATE $l \gets l +1$
\ENDWHILE
\RETURN ${\bmath \theta}$
\end{algorithmic}
\end{algorithm}

\section{Simulation Results}\label{simulations}
In this section, we provide results based on simulations to convince the robustness of our proposed calibration approach. We simulate an interferometric array with $R=47$ stations, with the longest baseline of about $30$ km. We simulate an observation centered at the north celestial pole (NCP), with a duration of $6$ hours at $150$ MHz. The integration time for each data sample is kept at $10$ s. For the full duration of the observation, there are $2160$ data points. Each data point consists of $1081$ baselines and $8$ real values corresponding to the $2\times 2$ complex visibility matrix.

\begin{figure}
\begin{minipage}{0.98\linewidth}
\centering
 \centerline{\epsfig{figure=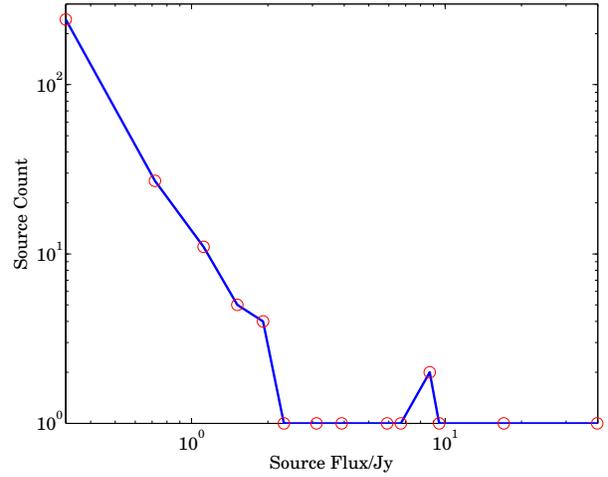,width=8.0cm}}
\end{minipage}
\caption{Histogram of the fluxes of the $300$ simulated sources. The peak flux is $40$ Jy.\label{flux_hist}}
\end{figure}

The sky is simulated to have $300$ sources, uniformly distributed over a field of view of $12\times 12$ degrees. The intensities of the sources are drawn using a power law distribution, with the peak intensity at $40$ Jy. In Fig. \ref{flux_hist}, we show the histogram of the intensities of the sources.
Our intention is to compare the fluxes of the weak sources, i.e. the sources with intensities less than or equal to $1$ Jy, after directional calibration is performed. In order to do that we corrupt the visibilities of the bright sources with directional errors that vary slowly with time. We consider three scenarios here: we only corrupt the signals of the sources that have intensities greater than (i) 1 Jy, (ii) 2 Jy and (iii) 5 Jy. For the simulated sky model, there are 28, 11 and 7 sources that have fluxes greater than 1 Jy, 2 Jy and 5 Jy, respectively. Note that in each case, we do not corrupt the signals of the weak sources as our only objective is to find the recovered flux after directional calibration and subtraction of the bright sources from the data, although in reality all sources will be corrupted by similar directional errors. Finally, we add zero mean white Gaussian noise to the simulated data, with the signal to noise ratio (SNR) defined as
\beq
\mathrm{SNR}\buildrel\triangle\over=10 \log_{10}\left(\frac{\sum_{p,q} \|{\bf {\sf V}}_{pq}\|^2}{\sum_{p,q} \| {\bf {\sf N}}_{pq}\|^2} \right)\ \ {\mathrm {dB}}.
\eeq
In all simulations, we have kept the SNR at $5$ dB.

\begin{figure}
\begin{minipage}{0.98\linewidth}
\centering
 \centerline{\epsfig{figure=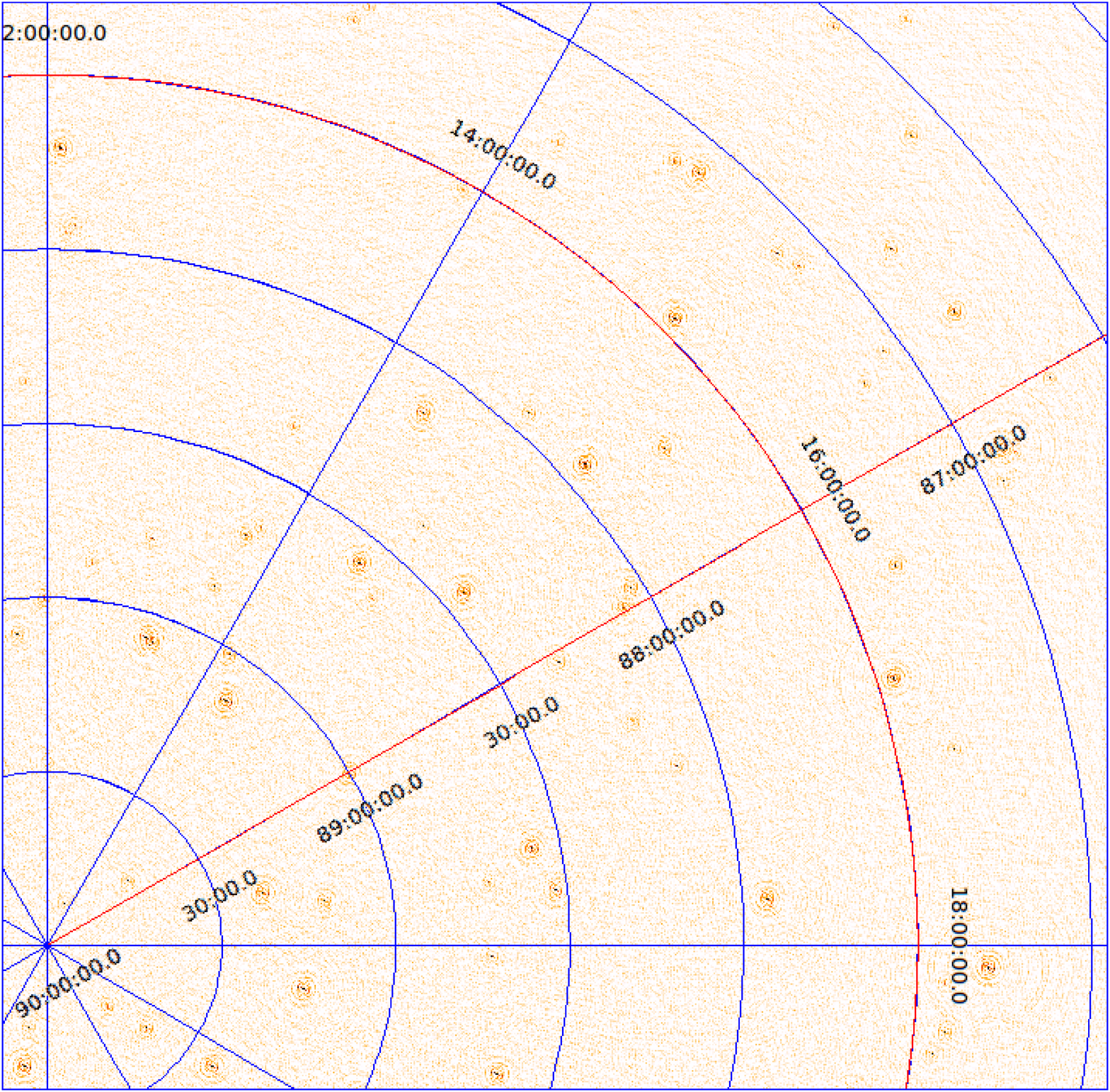,width=8.0cm}}
\end{minipage}
\caption{Simulated image of $4\times 4$ degrees of the sky, showing only weak sources with intensities less than 1 Jy.\label{img_original}}
\end{figure}

\begin{figure}
\begin{minipage}{0.98\linewidth}
\centering
 \centerline{\epsfig{figure=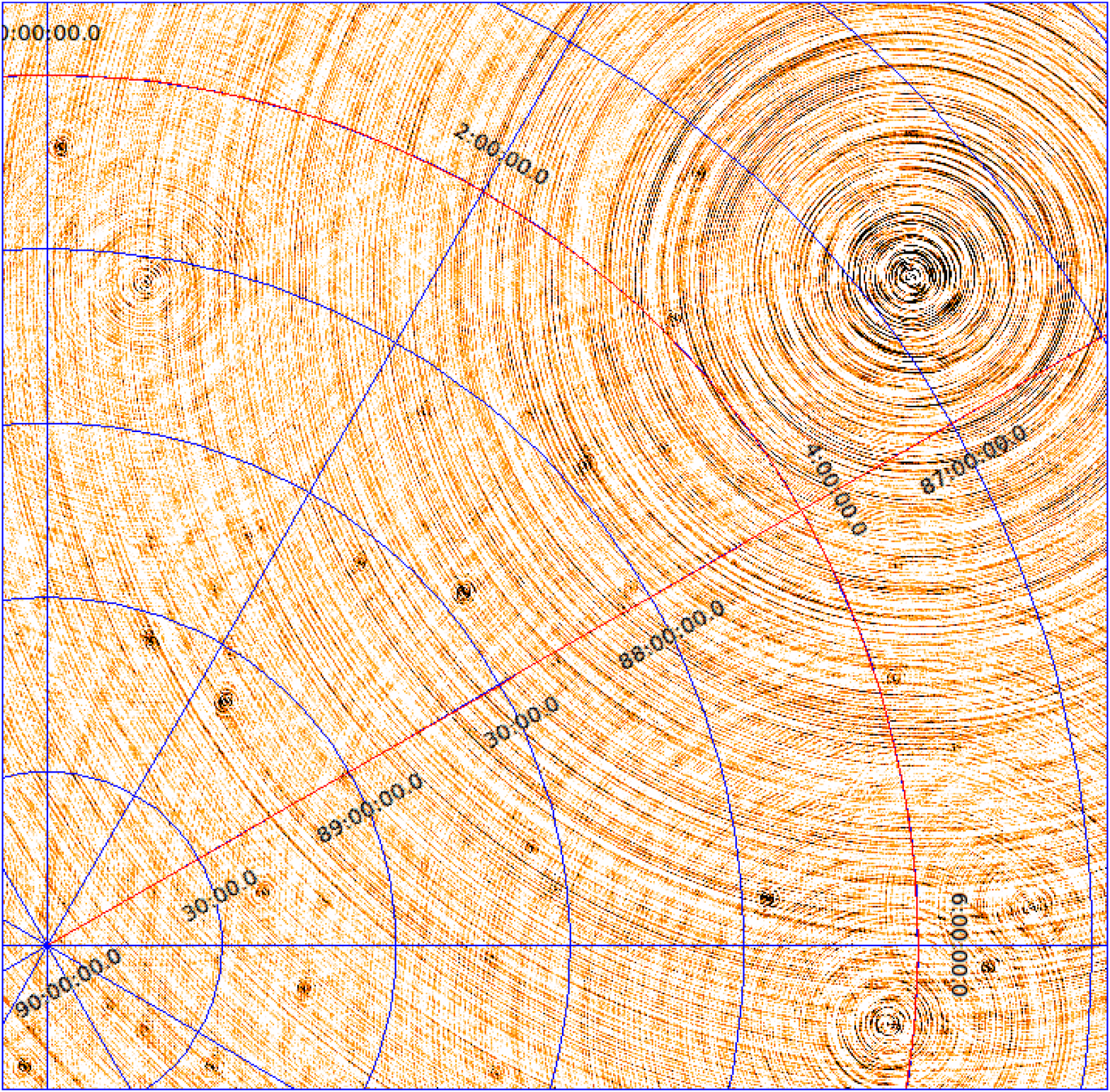,width=8.0cm}}
\end{minipage}
\caption{Simulated image of the sky where bright sources with fluxes greater than 1 Jy have been corrupted with directional errors. Due to these errors, there are artefacts throughout the image that makes it difficult to study the fainter background sources.\label{img_raw}}
\end{figure}

In Fig. \ref{img_original}, we show some of the weak sources (with intensities less than 1 Jy) over a $4\times 4$ degrees are of the field of view. In Fig. \ref{img_raw}, we have also added the bright sources with slowly varying directional errors. Note that in order to recover Fig. \ref{img_original} from Fig. \ref{img_raw}, directional calibration is essential.  

\begin{figure}
\begin{minipage}{0.98\linewidth}
\centering
 \centerline{\epsfig{figure=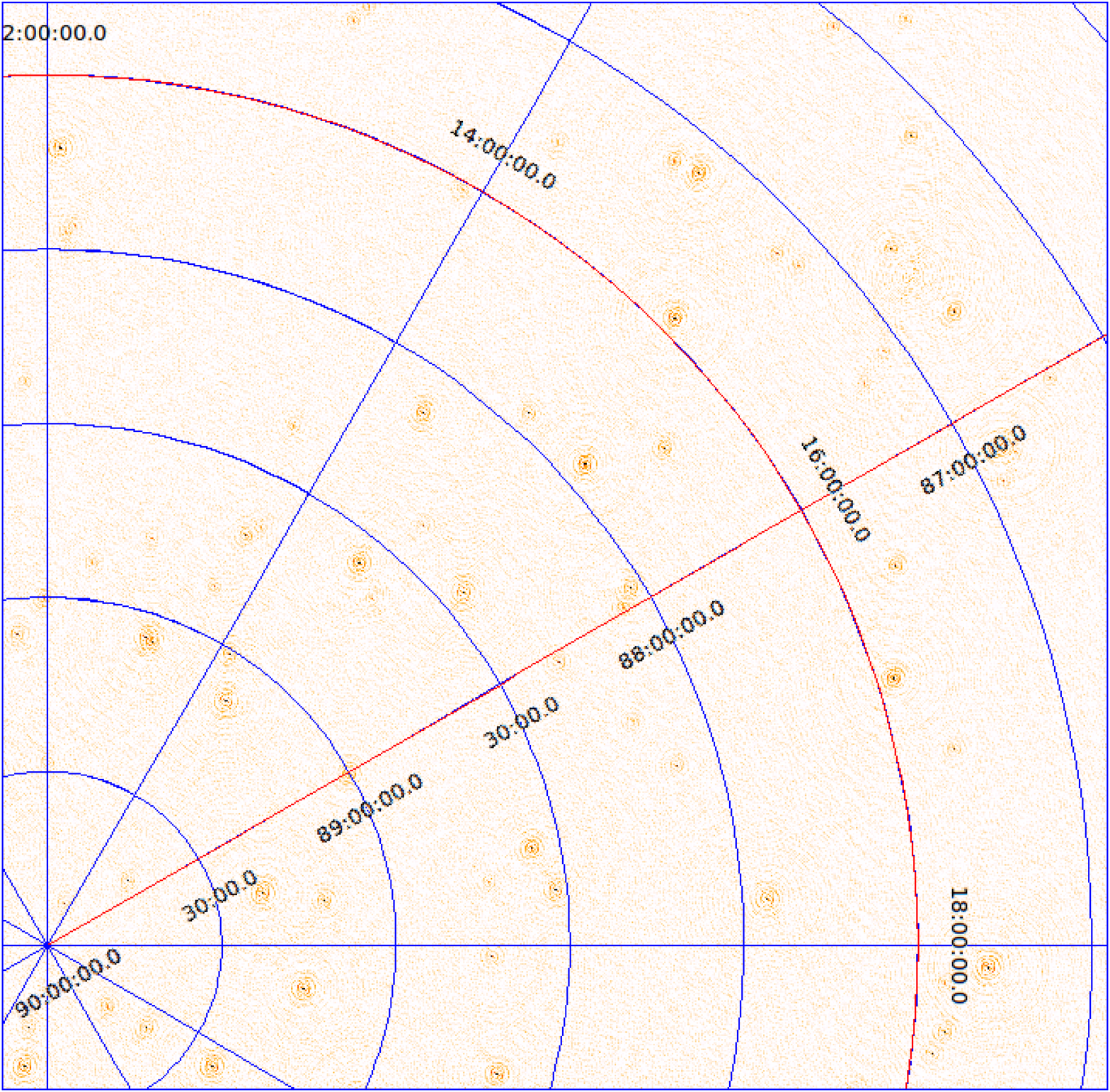,width=8.0cm}}
\end{minipage}
\caption{Image of the sky where the bright sources have been calibrated and subtracted from the data to reveal the fainter background sources. The traditional calibration based on a Gaussian noise model is applied.\label{img_normal}}
\end{figure}

\begin{figure}
\begin{minipage}{0.98\linewidth}
\centering
 \centerline{\epsfig{figure=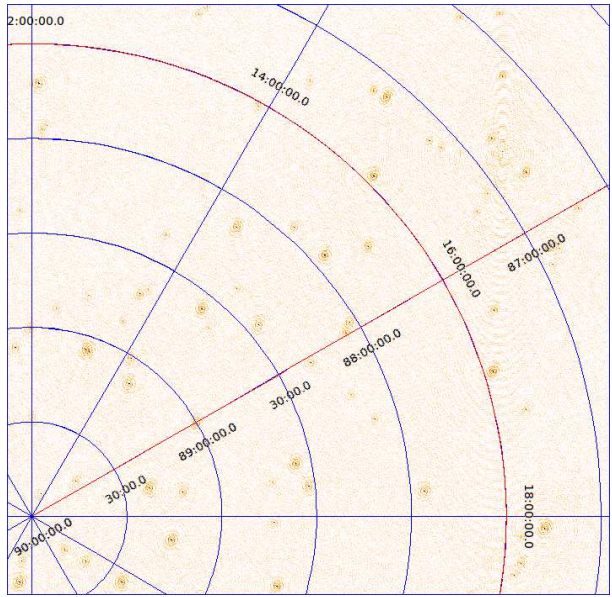,width=8.0cm}}
\end{minipage}
\caption{Image of the sky where the bright sources have been calibrated and subtracted from the data to reveal the fainter background sources. The robust calibration proposed in this paper is applied.\label{img_robust}}
\end{figure}

In Fig. \ref{img_normal}, we show the image after directional calibration along the bright sources and subtraction of their contribution from the data, using traditional calibration based on a Gaussian noise model. On the other hand, in Fig. \ref{img_robust}, we show the image after directional calibration and subtraction using a robust noise model. With respect to the subtraction of the bright sources from the data, both normal calibration and robust calibration show equal performance as seen from Figs. \ref{img_normal} and \ref{img_robust}.

We perform Monte Carlo simulations with different directional gain and additive noise realizations for each scenario (i), (ii) and (iii) as outlined previously. For each realization, we image the data after subtraction of the bright sources and compare the flux recovered for the weak sources before and after directional calibration. The directional calibration is performed for every $10$ time samples (every $100$ s duration). Therefore, the number of data points used for each calibration ($N$) is $10\times 1081\times 8=86480$ and the number of real parameters estimated are $47\times 8 \times 28=10528$, $47\times 8 \times 11=4136$ and $47\times 8 \times 7=2632$, respectively for scenarios (i),(ii) and (iii). For each scenario (i), (ii) and (iii), we perform 100 Monte Carlo simulations.

\begin{figure}
\begin{minipage}{0.98\linewidth}
\centering
 \centerline{\epsfig{figure=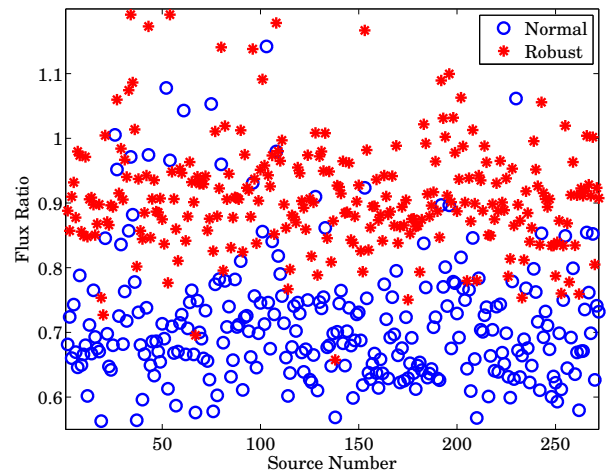,width=8.0cm}}
\end{minipage}
\caption{Ratio between the recovered flux and the original flux of each of the weak sources, when bright sources ($> 1$ Jy) are subtracted.\label{ratio_1}}
\end{figure}

\begin{figure}
\begin{minipage}{0.98\linewidth}
\centering
 \centerline{\epsfig{figure=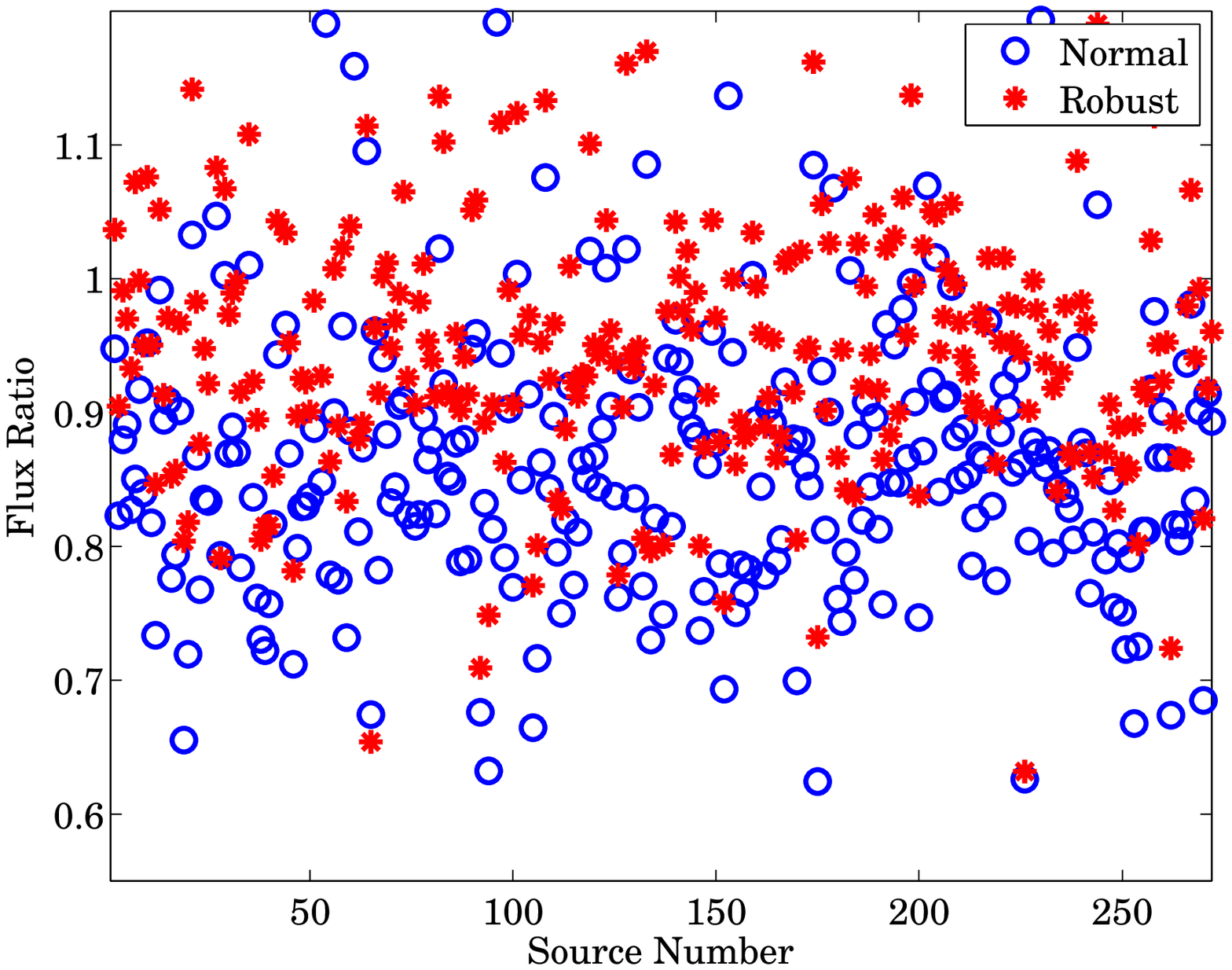,width=8.0cm}}
\end{minipage}
\caption{Ratio between the recovered flux and the original flux of each of the weak sources, when bright sources ($> 2$ Jy) are subtracted.\label{ratio_2}}
\end{figure}

\begin{figure}
\begin{minipage}{0.98\linewidth}
\centering
 \centerline{\epsfig{figure=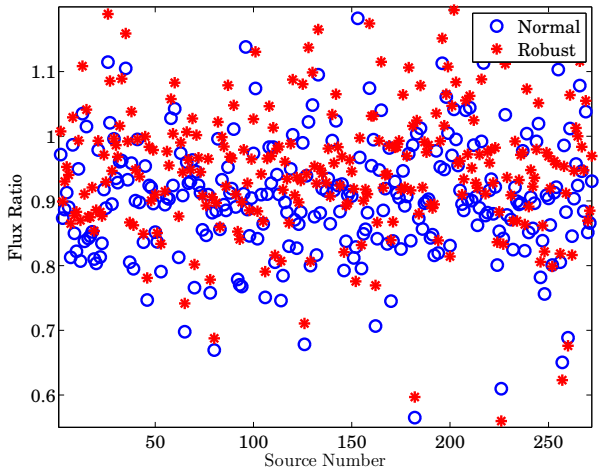,width=8.0cm}}
\end{minipage}
\caption{Ratio between the recovered flux and the original flux of each of the weak sources, when bright sources ($> 5$ Jy) are subtracted.\label{ratio_5}}
\end{figure}

Our performance metric is the ratio between the recovered peak flux of the weak sources compared to the original flux of each source. We calculate the average ratio (recovered flux / original flux) over all Monte Carlo simulations. In Figs. \ref{ratio_1}, \ref{ratio_2} and \ref{ratio_5}, we show the results obtained for scenario (i),(ii) and (iii), respectively.

\begin{figure}
\begin{minipage}{0.98\linewidth}
\centering
 \centerline{\epsfig{figure=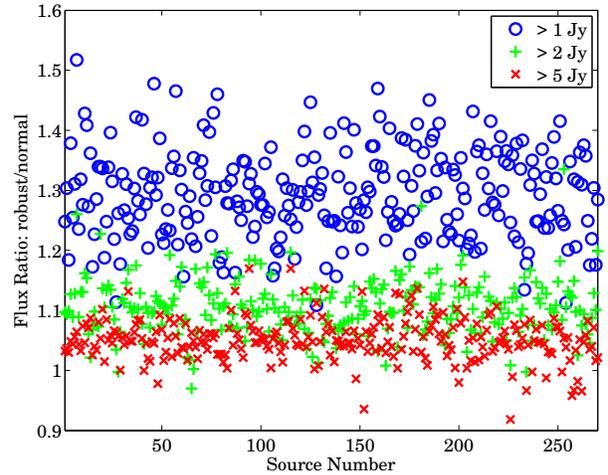,width=8.0cm}}
\end{minipage}
\caption{Ratio between the recovered flux using robust calibration and the recovered flux using normal calibration. Almost always, robust calibration recovers more flux compared with normal calibration. The different colours indicate different scenarios where the number of bright sources subtracted is varied.\label{ratio_rn}}
\end{figure}

We observe two major characteristics in Figs.  \ref{ratio_1}, \ref{ratio_2} and \ref{ratio_5}. First, we see that as we calibrate over an increasing number of directions (and subtract an increasing number of sources), the recovered flux is reduced. Second, in all scenarios, robust calibration recovers more flux compared to normal calibration. To illustrate this point, we also plot in Fig. \ref{ratio_rn}, the ratio between the recovered flux using robust calibration and the recovered flux using normal calibration. As we see from Fig. \ref{ratio_rn}, we almost always get a value  greater than 1 for this ratio, indicating that we recover more flux using robust calibration.

We summarize our findings in Table \ref{comptab}. We see that at worst case, the performance of normal calibration gives a flux reduction of about 20\% compared to robust calibration.
\setcounter{table}{0}
\begin{table*}
\centering
\begin{tabular}{cccc}\hline
\multicolumn{1}{c}{\multirow{4}{1in}{No. of sources calibrated and subtracted } }& 
\multicolumn{1}{c}{\multirow{4}{1in}{Lowest flux of the subtracted sources (Jy)} }& 
\multicolumn{2}{c}{\multirow{1}{2in}{Average reduction of the flux of weak background sources (\%)}} \\ 
\\
\\
\\
& & \multicolumn{1}{c}{Normal Calibration} &\multicolumn{1}{c}{Robust Calibration} \\\hline
 28 & 1 &  28.7 & 8.2 \\\hline
 11 & 2 & 12.3 & 3.2 \\\hline
 7 & 5 & 7.9 & 3.1 \\\hline
\end{tabular}
\caption{Comparison of the reduction of flux of the weak background sources with normal and robust calibration.}\label{comptab}
\end{table*}

Up to now, we have only considered the sky to consist of only point sources. In reality, there is diffuse structure in the sky. This diffuse structure is seldom incorporated into the sky model during calibration either because it is too faint or because of the complexity of modeling it accurately. We have also done simulations where there is faint diffuse structure in the sky and only the bright foreground sources are calibrated and subtracted. We have chosen scenario (i) in the previous simulation except that we have replaced the sources below $1$ Jy with Gaussian sources with peak intensities below $1$ Jy and with random shapes and orientations. 

\begin{figure}
\begin{minipage}{0.98\linewidth}
\centering
 \centerline{\epsfig{figure=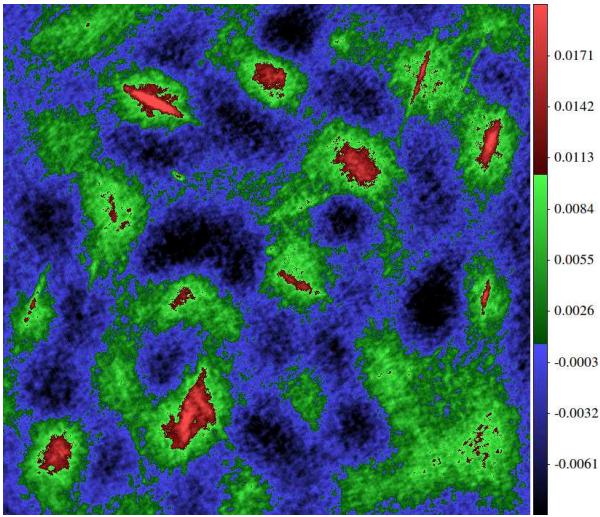,width=8.0cm}}
\end{minipage}
\caption{Average residual image of the diffuse structure after subtracting the bright sources by normal calibration. The colour scale is in Jy/PSF.\label{nor_gauss}}
\end{figure}

\begin{figure}
\begin{minipage}{0.98\linewidth}
\centering
 \centerline{\epsfig{figure=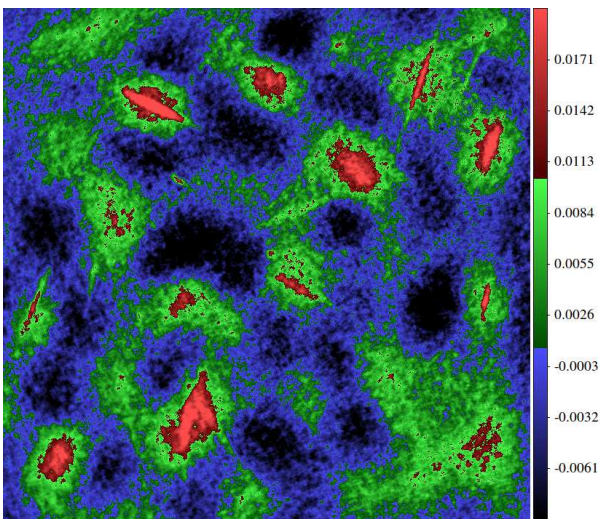,width=8.0cm}}
\end{minipage}
\caption{Average residual image of the diffuse structure after subtracting the bright sources by robust calibration. The colour scale is in Jy/PSF and is the same as in Fig. \ref{nor_gauss}.\label{robust_gauss}}
\end{figure}

In Fig. \ref{nor_gauss}, we have shown the residual image of a $6\times 6$ degrees area in the sky after removing all sources brighter than $1$ Jy. The residual image is obtained by averaging $100$ Monte Carlo simulations. The equivalent image for robust calibration is given in Fig. \ref{robust_gauss}.

As seen from Figs. \ref{nor_gauss} and \ref{robust_gauss}, there is more flux in the diffuse structure after robust calibration. This is clearly seen in the bottom right hand corner of both figures where Fig. \ref{robust_gauss} has more positive flux than in Fig. \ref{nor_gauss}.

\section{Conclusions}\label{conclusions}
We have presented the use of Student's t distribution in radio interferometric calibration. Compared with traditional calibration that has an underlying Gaussian noise model, robust calibration using Student's t distribution can handle situations where there are model errors or outliers in the data. Moreover, by automatically selecting the number of degrees of freedom $\nu$ during calibration, we have the flexibility of choosing the appropriate distribution even when no outliers are present and the noise is perfectly Gaussian.  For the specific case considered in this paper, i.e. the loss of coherency or flux of unmodeled sources, we have given simulation results that show the significantly improved flux preservation with robust calibration. Future work would focus on adopting this for pipeline processing of massive datasets from new and upcoming radio telescopes. 

\section*{Acknowledgments}
We thank the reviewer, Fred Schwab, for his careful review and insightful comments. We also thank the Editor and Assistant Editor for their comments. We also thank Wim Brouw for commenting on an earlier version of this paper.
\bibliographystyle{mn2e}
\bibliography{references}

\bsp
\label{lastpage}
\end{document}